# Neural ranking models for document retrieval

Mohamed Trabelsi[1] · Zhiyu Chen[1] · Brian D. Davison[1] · Jeff Heflin[1]

**Abstract**
Ranking models are the main components of information retrieval systems. Several approaches to ranking are based on traditional machine learning algorithms using a set of hand-crafted features. Recently, researchers have leveraged deep learning models in information retrieval. These models are trained end-to-end to extract features from the raw data for ranking tasks, so that they overcome the limitations of hand-crafted features. A variety of deep learning models have been proposed, and each model presents a set of neural network components to extract features that are used for ranking. In this paper, we compare the proposed models in the literature along different dimensions in order to understand the major contributions and limitations of each model. In our discussion of the literature, we analyze the promising neural components, and propose future research directions. We also show the analogy between document retrieval and other retrieval tasks where the items to be ranked are structured documents, answers, images and videos.

## 1 Introduction

Recent advances in neural networks enable the improvement in the performance of multiple fields including computer vision, natural language processing, machine translation, speech recognition, etc. The main neural components that led to the breakthrough in multiple fields are convolutional and recurrent neural networks. Information retrieval (IR) also benefits from deep neural network models leading to state-of-the-art results in multiple tasks.

✉ Mohamed Trabelsi
  mot218@lehigh.edu

  Zhiyu Chen
  zhc415@lehigh.edu

  Brian D. Davison
  davison@cse.lehigh.edu

  Jeff Heflin
  heflin@cse.lehigh.edu

[1] Computer Science and Engineering, Lehigh University, Bethlehem, PA, USA

Retrieval models take as input a user's query, and then present a set of documents that are relevant to the query. In order to return a useful set of documents to the user, the retrieval model should be able to rank documents based on the given query. This means that the model ranks the documents using features from both the query and documents. Traditional ranking models for text data might utilize OKAPI/BM25 (Robertson et al., 1994) which computes the score of matching between the query and document based in part on the presence of query terms in each document. Machine learning algorithms can learn ranking models, and the input to these models are a set of often hand-crafted features. This setting is known as learning to rank (LTR) using hand-crafted features. These features are domain specific and time-consuming in terms of defining, extracting, and validating a set of specific features for a given task. In order to overcome the limitations of using hand-crafted features, researchers proposed deep ranking models that accept raw text data as an input and learn suitable representations for inputs and ranking functions.

A key feature in information retrieval models is the relevance judgement. A ranking model with sufficient capacity is needed to capture the matching signals, and map document-query pairs to accurate prediction of a real-valued relevance score. Deep neural networks are known for their ability to capture complex patterns in both feature extraction and model building phases. Due to the advantages of deep neural networks, researchers have focused on designing neural ranking models to learn both features and model simultaneously.

Neural ranking models have many challenges to address in information retrieval tasks. First, the queries and documents have different lengths: the query is usually a short text that typically consists of a few keywords, and the document is long with both relevant and irrelevant parts to the query. Second, in many cases, the query and documents have different terms, so exact matching models cannot be used to accurately rank documents; a neural matching model should be designed to capture semantic matching signals to predict the relevance score. The semantic similarity is context dependent, and another challenge for the neural ranking model is to understand the context of both query and documents in order to generalize across multiple domains.

Many neural ranking models have been proposed primarily to solve information retrieval tasks. Other neural models are proposed for text matching tasks, and they are used in ad-hoc retrieval because understanding the semantic similarity between sentences in text matching can enhance retrieval results mainly for sentence or passage level document retrieval scenarios. So, in addition to the neural ranking models that are introduced specifically for retrieval tasks, we will review multiple text matching-based neural models that can be applied to document retrieval.

Existing surveys on neural ranking models focus on the embedding layer that maps tokens to embedding vectors known as word embeddings. Onal et al. (2017) classified existing publications based on the IR tasks. For each task, the authors discussed how to integrate word embeddings in neural ranking models. In particular, the authors proposed two categories based on how the word embedding is used. For the first category, the neural ranking models use a pre-trained word embedding to aggregate embeddings with average or sum of word embeddings, or to compute cosine similarities between word embeddings. The second category includes end-to-end neural ranking models where the word embedding is learned or updated while training the neural ranking model. Mitra and Craswell (2018) presented a tutorial for document retrieval with a focus on traditional word embedding techniques such as Latent Semantic Analysis (LSA) (Deerwester et al., 1990), word2vec (Mikolov et al., 2013a), Glove (Pennington et al., 2014a), and paragraph2vec (Le & Mikolov, 2014). The authors reviewed multiple neural toolkits as part of the tutorial and a few deep neural models for IR. Guo et al.

(2019) reviewed the learning strategies and the major architectures of neural ranking models. Lin et al. (2020) focused mainly on pretrained Transformers (Vaswani et al., 2017) for text ranking, where they showed that a BERT-based multi-stage ranking model is a potential choice for a tradeoff between effectiveness and efficiency of a neural ranking model. Xu et al. (2020b) reviewed deep learning models for matching in document retrieval and recommendation systems. The authors grouped the neural ranking models into two categories which are representation learning and matching function learning. Compared to the survey of Xu et al. (2020b), in addition to grouping neural ranking models into five categories based on the neural components and design (Sect. 6), we summarize multiple models based on nine features that are frequently presented in the neural ranking models (Sect. 7). We also discuss the generalization of neural ranking models to other retrieval tasks with different objects to query and rank. In particular, we describe four retrieval tasks which are: structured document retrieval, Question-Answering, image retrieval, and Ad-hoc video search (Sect. 8).

In conclusion, the objective of our survey is to summarize the current progress, and compare multiple neural architectures using different dimensions. Our comparison is more fine-grained than existing surveys in terms of grouping and decomposing neural ranking models into important neural components and architecture designs. In addition, our survey includes an overview of models in the literature based on several features that are frequently presented in neural ranking models. The detailed comparison of multiple neural ranking models can help researchers to identify the common neural components that are used in the document retrieval task, understand the main benefits from using a given neural component, and investigate the promising neural components in future research to improve document retrieval results.

In summary, the content of this survey paper is organized as follows:

- In Sect. 2, we briefly introduce the deep learning terminologies and techniques to which we refer when discussing the different neural ranking models in the literature.
- In Sect. 3, we introduce the document retrieval task, and present the flowchart of the neural ranking based models for document retrieval.
- In Sect. 4, we formulate the document retrieval task using the LTR framework.
- In Sect. 5, we review the three LTR approaches used in ad-hoc retrieval namely, pointwise, pairwise, and listwise approaches.
- In Sect. 6, we summarize the neural ranking models in ad-hoc retrieval, and we propose placing the neural ranking models into five groups, where each group has unique neural components and architectures.
- In Sect. 7, we discuss the features of neural ranking models, and propose nine features that are presented in neural ranking models. These features are used for an in-depth overview of several neural ranking models that are proposed in the literature.
- In Sect. 8, we show that neural ranking models can generalize beyond document retrieval. In particular, we describe four related retrieval tasks: structured document retrieval, question-answering, image retrieval, and ad-hoc video search.
- In Sect. 9, we summarize the current best practices for the design of neural ranking models, and we discuss multiple research directions in document retrieval.

## 2 Background

In this section, we briefly introduce the deep learning terminology and techniques most commonly used in ad-hoc retrieval. This includes Convolutional Neural Networks (CNN) (LeCun & Bengio, 1998), Recurrent Neural Networks (RNN) (Elman, 1990), Long Short-Term Memory (LSTM) (Hochreiter & Schmidhuber, 1997), Gated Recurrent Units (GRU) (Cho et al., 2014), word embedding techniques (Wang et al., 2020), attention mechanism (Bahdanau et al., 2015), deep contextualized language models (Peters et al., 2018; Devlin et al., 2019), and knowledge graphs (Wang et al., 2017). We will be referring to these neural components and techniques when discussing the different neural ranking models in the literature.

### 2.1 Convolutional neural network (CNN)

A CNN (LeCun & Bengio, 1998) extracts features from data by defining a set of filters or kernels that spatially connect local regions. Compared to the dense networks, each neuron is connected to only a small number of neurons instead of being connected to all neurons from the previous layer. This design significantly reduces the number of parameters in the model. In addition, the weights in CNN filters are shared among multiple local regions for the input, and this further reduces the number of parameters. The outputs of the CNN are called feature maps. Pooling operations, such as average and max pooling, are usually applied to the feature map to keep only the significant signals, and to further reduce the dimensionality. A CNN kernel has a predefined size so that in order to handle the information in the border of the input, padding is introduced to enlarge the input. CNNs were first introduced to solve image-related tasks such as image classification (Krizhevsky et al., 2017; Simonyan & Zisserman, 2015; Szegedy et al., 2015; Ioffe & Szegedy, 2015; Szegedy et al., 2016), and were later adapted to solve text-related tasks such as NLP and information retrieval (Collobert et al., 2011; Dai et al., 2018; Hui et al., 2017; Jaech et al., 2017; Lan & Xu, 2018; McDonald et al., 2018; Pang et al., 2016b; Tang & Yang, 2019; Kalchbrenner et al., 2014).

### 2.2 Recurrent neural network (RNN)

An RNN (Elman, 1990) learns features and long-term dependencies from sequential and time-series data. RNN reads the input sequence sequentially to produce a hidden state in each timestamp. These hidden states can be seen as memory-cells that store the sequence information. A current hidden state is a function of the previous hidden state and the current input. Therefore, a hidden state is computed for each timestamp, and the hidden state that corresponds to the last timestamp in the sequence captures the context-aware representation of the entire sequence. Two major problems of the vanilla RNN are the vanishing and exploding gradients (Pascanu et al., 2013; Sutskever et al., 2011) that can occur after back-propagation through time during the training phase. For example, for long sequences, when the gradient flows from later to earlier timestamps in the input sequence, the signal from the gradient can become very weak or vanish. Two variants of RNN which are LSTM and GRU have been proposed to capture long-term dependencies better than RNN, and therefore reduce the vanishing and exploding of the gradient. The new structures of LSTM and GRU allow the network to capture long-range dependencies.

## 2.3 Long short-term memory (LSTM)

Exploding and vanishing gradients during the training phase result in the failure of the network to learn long-term dependencies in the input data. LSTM (Hochreiter & Schmidhuber, 1997) was introduced to mitigate the effects of exploding and vanishing gradients. LSTM differs from the vanilla RNN in the structure of the memory cell where three gates are introduced to control the information in the memory cell. First, the input gate controls the influence of the current input on the memory cell. Second, the forget gate controls the previous information that should be forgotten in the current timestamp. Third, the output gate controls the influence of the memory cell on the hidden state of the current timestamp. Compared to RNN, LSTM has led to significant improvements in multiple fields with sequential data such as text, video, and speech. LSTM has been applied to solve multiple tasks including language modeling (Kim et al., 2016), text classification (Dai & Le, 2015), machine translation (Sutskever et al., 2014), video analysis (Singh et al., 2017), image captioning (Karpathy & Fei-Fei, 2017), and speech recognition (Graves et al., 2013).

## 2.4 Gated recurrent units (GRU)

Similar to LSTM, GRU (Cho et al., 2014) is used for sequence-based tasks to capture long-term dependencies. However, unlike LSTM, GRU does not have separate memory cells. In LSTM, the output gate is used to control the memory content that is used by other units in the network. On the other hand, the GRU model does not contain an output gate, and therefore uses its content without any gating control. In addition, while LSTM computes the value of the new added memory independently of the forget gate, GRU does not independently control the new added activation but uses the reset gate to control the previous hidden state. More differences and similarities between LSTM and GRU are summarized by Chung et al., (2014). This model has been shown to achieve good performance in multiple tasks such as machine translation (Cho et al., 2014) and sentiment classification (Tang et al., 2015) .

## 2.5 Attention mechanism

The attention mechanism was first proposed by Bahdanau et al., (2015) for neural machine translation. The original Seq2Seq model (Sutskever et al., 2014) used an LSTM to encode a sentence from its source language and another LSTM to decode the sentence into a target language. However, this approach was unable to capture long-term dependencies. In order to solve this problem, Bahdanau et al., (2015) proposed to simultaneously learn to align and translate the text. They learn attention weights which can produce context vectors that focus on a set of positions in a source sentence when predicting a target word. The attention vector is computed using a weighted sum of all the hidden states of an input sequence, where a given attention weight indicates the importance of a token from the source sequence in the attention vector of a token from the output sequence. Although introduced for machine translation, the attention mechanism has been a useful tool in many tasks such as document retrieval (McDonald et al., 2018), document classification (Yang et al., 2016b), sentiment classification (Wang et al., 2017a), recommender systems (Ying et al., 2018), speech recognition (Chan et al., 2016), and visual question answering (Lu et al., 2016).

## 2.6 Word embedding

Words are embedded into low dimensional real-valued vectors based on the distributional hypothesis (Harris, 1954). In many proposed models, the context is defined as the words that precede and follow a given target word in a fixed window (Bengio et al., 2003; Mnih & Hinton, 2007; Mikolov et al., 2013a; Pennington et al., 2014b). Mikolov et al., (2013b) proposed the Skip-gram model which scales to a corpora with billions of words. These pre-trained word embeddings are a key component in multiple models in neural language understanding and information retrieval tasks. However, there are multiple challenges for learning word embeddings. First, the embedding should capture the syntax and semantics of tokens. Second, the embedding should model the polysemy characteristic where a given word can have different meanings depending on the context. Multiple works have been proposed to produce context-sensitive embeddings for tokens that capture not only the meaning of a token, but also the contextual information of a token. Researchers have investigated the use of RNN to produce context-dependent representations for tokens (Yang et al., 2017; Ma & Hovy, 2016; Lample et al., 2016). The word embeddings are initialized using pre-trained embeddings, then the parameters of RNN are learned using labeled data from a given task. Peters et al., (2017) proposed a semi-supervised approach to train a context-sensitive embedding using a neural language model pre-trained on a large and unlabeled corpus. A forward and backward RNN are used to predict the next token so that the neural language model encodes both the semantic and syntactic features of tokens in a context. Adding pre-trained context-sensitive embeddings from both the forward and backward RNN improves the performance of the sequence tagging task. However, the embeddings are not deep, in the sense that they are not a function of all of the internal layers of both the forward and background language models. Recently, researchers have focused on creating deep context-sensitive embeddings such as ELMo (Peters et al., 2018) and BERT (Devlin et al., 2019) which are trained on large amounts of unlabeled data and achieve high performance in multiple NLP tasks.

## 2.7 Deep contextualized language models

Peters et al. (2018) proposed ELMo which is a deep contextualized language model composed of forward and backward LSTMs. ELMo produces deep embeddings, where the representations from multiple layers of both the forward and backward LSTM are linearly combined using task-specific learnable weights to compute the embedding of a given token. Combining the internal states leads to richer embeddings. Although ELMo improves the results of multiple NLP tasks, ELMo decouples the left-to-right and right-to-left contexts by using a shallow concatenation of internal states from independently trained forward and backward LSTMs. Devlin et al. (2019) proposed a language model, called Bidirectional Encoder Representations from Transformers (BERT), that fuses both left and right contexts.

BERT is a deep contextualized language model that contains multiple layers of Transformer (Vaswani et al., 2017) blocks. Each block has a multi-head self-attention structure followed by a feed-forward network, and it outputs contextualized embeddings for each token in the input. BERT is trained on large collections of unlabeled data over two pretraining tasks which are next sentence prediction and the masked language model. After the pre-training phase, BERT can be used for downstream tasks on single text or text pairs using special tokens ([SEP] and [CLS]) that are added into the input. For single text classification, [CLS] and [SEP] are added to the beginning and the end of the sequence, respectively. For text pairs-based applications, BERT encodes the text pairs using bidirectional cross attention between the two sentences. In this case, the text pair is concatenated using [SEP], and then BERT treats the concatenated text as a single text. The sentence pair classification setting is used to solve multiple tasks in information retrieval including document retrieval (Dai & Callan, 2019; Nogueira et al., 2019; Yang et al., 2019a), passage re-ranking (Nogueira & Cho, 2019), frequently asked question retrieval (Sakata et al., 2019), table retrieval (Chen et al., 2020b), and semantic labeling (Trabelsi et al., 2020a). The single sentence setting is used for text classification (Sun et al., 2019; Yu et al., 2019). BERT takes the final hidden state $\mathbf{h}_\theta$ of the first token [CLS] as the representation of the whole input sequence, where $\theta$ denotes the parameters of BERT. Then, a simple softmax layer, with parameters $W$, is added on top of BERT to predict the probability of a given label $l$:

$$p(l \mid \mathbf{h}_\theta) = \text{softmax}(W\mathbf{h}_\theta) \quad (1)$$

The parameters of BERT, denoted by $\theta$, and the softmax layer parameters $W$ are fine-tuned by maximizing the log-probability of the true label.

### 2.8 Knowledge graphs

Large scale general domain knowledge bases (KB), like Freebase (Bollacker et al., 2008) and DBpedia (Lehmann et al., 2015), contain rich semantics that can be used to improve the results of multiple tasks in natural language processing and information retrieval. Knowledge bases contain human knowledge about entities, classes, relations, and descriptions. Knowledge can be represented as graphs and this leads to the concept of knowledge graphs (KG). Various methods have been proposed for representation learning of knowledge graphs, which aims to project entities and relations into a continuous space. TransE (Bordes et al., 2013), inspired by Word2Vec (Mikolov et al., 2013b), is the most representative translation-based model, which considers the translation operation between head and tail entities for relations. The variants of TransE, such as TransH (Wang et al., 2014) and TransR (Lin et al., 2015), follow a similar principle but use different scoring functions to learn the embeddings. Socher et al. (2013a) apply neural tensor networks to learn knowledge graph embeddings. Dettmers et al. (2018) propose a convolutional neural network approach to learn knowledge graph embeddings and use them to perform link prediction. RDF2Vec (Ristoski & Paulheim, 2016) adapts the Word2Vec (Mikolov et al., 2013b) approach to RDF graphs in order to learn embeddings for entities in RDF graphs.

Recently, researchers have explored a new direction called graph neural networks (Cai et al., 2017) to solve multiple tasks (Zhou et al., 2018). Graph neural networks capture rich relational structures between nodes in a graph using message passing and encode the global structure of a graph in low dimensional feature vectors known as graph embeddings. The Graph Convolutional Network (GCN) (Kipf & Welling, 2017) can capture high order neighborhood information to learn representations of nodes in a graph. The success of GNNs has boosted research on various tasks. R-GCN (Schlichtkrull et al. 2018) pioneered the use of graph convolutional networks to model relations in knowledge graphs. The embeddings learned by R-GCN have shown to be effective for downstream tasks such as entity classification and link prediction. More recently, Xu et al. (2020a) first construct a product knowledge graph and then propose a self-attention-enhanced distributed representation learning method with an efficient multi-task training schema to learn the graph embeddings, which can improve the performance of downstream tasks such as search ranking and recommendation. GNNs are also introduced in table retrieval, where Trabelsi et al. (2020c) first construct a knowledge graph from table collections and then learn the graph embeddings in order to rank tables.

## 3 Document retrieval

Information retrieval is a broad research area that covers a variety of content types and tasks. In this survey, we focus on the document retrieval and ranking task and propose detailed descriptions and groupings of multiple neural ranking models in the document retrieval. In terms of scope, we focus on text-based document retrieval, where the input to the neural ranking model is raw text, and the output is a ranked list of documents. There are two advantages from retrieving documents using text-based neural ranking models. First, the textual input form for the query and document can be used directly by the neural ranking model, so that text-based ranking models generalize better than traditional handcrafted models which need specific features. Second, text-based ranking models provide additional signals such as semantic and relevance matching signals to accurately predict the relevance score.

Ranking and retrieving documents that are relevant to a user's query is a classic information retrieval task. Given a query, the ranking model outputs a ranked list of documents so that the top ranked items should be more relevant to the user's query. Search engines are examples of systems that implement ad-hoc retrieval where the possible number of queries that are continually submitted to the system is huge. The general flowchart of document retrieval with neural ranking models is illustrated in Fig. 1. A large collection of documents is indexed for a fast retrieval. A user enters a text-based query which goes through a query processing step consisting of a query reformulation and expansion (Azad & Deepak, 2019). Many neural ranking models have complex architectures, therefore computing the query-document relevance score using the neural ranking model for every document in the initial large collection of documents leads to a significant increase in the latency for obtaining a ranked list of documents from the user's side. So, the neural ranking component is usually used as a re-ranking step that takes two inputs which are the candidate documents and the processed query. The candidate documents are obtained from an unsupervised ranking stage, such as BM25, which takes as inputs the initial set of indexed documents and the processed query. During the unsupervised ranking, recall is more important than precision to cover all possible relevant documents and forward a set of candidate documents, that has both relevant and irrelevant documents, to the neural based re-ranking stage. The output of the ranking model is a set of relevant documents to the user's query which are returned to the user in a particular order.

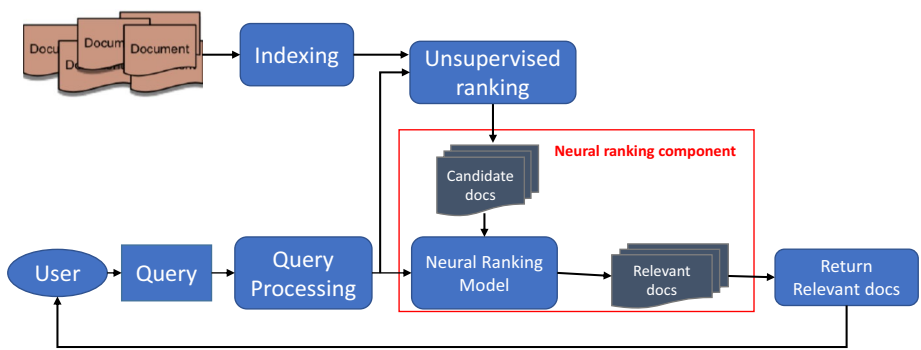

**Fig. 1** Overview of the flowchart of the neural ranking based document retrieval. The neural ranking component is highlighted within the red box. The inputs to the neural ranking model are the processed query and the candidate documents that are obtained from the traditional ranking phase. The final output of the neural ranking model is a ranking of relevant documents to the user's query

The inputs to neural ranking models consist of queries and documents with variable lengths in which the ranking model usually faces a short query with keywords, and long documents from different authors with a large vocabulary. Although exact matching is an important signal in retrieval tasks, ranking models also need to semantically match queries and documents in order to accommodate vocabulary mismatch. In ad-hoc retrieval, features can be extracted from documents, queries, and document-query interactions. Some document features go beyond text content and can include number of incoming links, page rank, metadata, etc. A challenging scenario for a ranking model is to predict the relevance score by only using the document's textual content, because there is no guarantee to have additional features when ranking documents. Neural ranking models have been used to extract feature representations for query and document using text data. For example, a deep neural network model can be used to map the query and documents to feature vectors independently, and then a relevance score is calculated using the extracted features. For query-document interaction, classic information retrieval models like BM25 can be considered as a query-document feature. For neural ranking models with a textual input for query and document, features are extracted from the local interactions between query and document.

## 4 Task formulation

For ranking tasks, the objective is to output a ranked list of documents given a query representing an information need. Neural ranking models are trained using the LTR framework. Thus, here we present the LTR formulation for retrieval tasks.

The LTR framework starts with a phase to train a model to predict the relevance score from a given query-document pair. During the training phase, a set of queries $Q = \{q^1, q^2, \ldots, q^{|Q|}\}$ and a large collection of documents $\mathcal{D}$ are provided. Without loss of generality, we suppose that the number of tokens in a given query is $n$, and the number of tokens in a given document is $m$. The groundtruth relevance scores for query-document pairs are needed to train the neural ranking model. In the general setting, for a given query, the groundtruth relevance scores are only known for a subset of documents from the large collection of documents $\mathcal{D}$. So, we formally define that each query $q^i$ is associated with a subset of documents $d^i = (d^i_1, d^i_2, \ldots, d^i_{l^i})$ from $\mathcal{D}$, where $d^i_j$ denotes the $j^{th}$ document for the $i^{th}$ query, and $l^i$ is the size of $d^i$. Each list of documents $d^i$ is associated with a list of relevance scores $y^i = (y^i_1, y^i_2, \ldots, y^i_{l^i})$ where $y^i_j$ denotes the relevance score of document $d^i_j$ with respect to query $q^i$. The objective is to train a function $f_w$, with parameters $w$, that is used to predict the relevance score $z^i_j = f_w(q^i, d^i_j)$ of a given query-document pair $(q^i, d^i_j)$.

The function $f_w$ is trained by minimizing a loss function $L(w)$. In LTR, the learning categories are grouped into three groups based on their training objectives: the pointwise, pairwise, and listwise approaches. In the next section, we will briefly describe these three learning categories.

In general, $f_w$ is considered as the composition of two functions $M$ and $F$, with $F$ is a feature extractor function, and $M$ is a ranking model. So for a given query-document pair $(q^i, d_j^i)$, $z_j^i$ is given by:

$$z_j^i = f_w(q^i, d_j^i) = M \circ F(q^i, d_j^i) \qquad (2)$$

In traditional ranking models, the function $F$ represents a set of hand-crafted features. The set of hand-crafted features include query, document, and query-document features. A ranking model $M$ is trained to map the feature vector $F(q^i, d_j^i)$ into a real-valued relevance score such that the most relevant documents to a given query are scored higher to maximize a rank-based metric.

In recently proposed ranking models, deep learning architectures are leveraged to learn both feature vectors and models simultaneously. The features are extracted from query, document, and query-document interactions. The neural ranking models are trained using ground truth relevance of query-document pairs. The main objective of this article is to discuss the deep neural architectures that are proposed for the document retrieval task. To describe the overall steps of training neural ranking models, in the next section, we give a brief overview about the different learning strategies before presenting the existing neural ranking models.

## 5 Categories of strategies for learning for ad-hoc retrieval

Liu (2009) divided LTR approaches into three categories based on their training objectives. In the pointwise category, each query-document pair is associated with a real-valued relevance score, and the objective of the training is to make a prediction of the exact relevance score using existing classification (Gey, 1994; Li et al., 2007) or regression models (Cossock & Zhang, 2006; Fuhr, 1989). However, predicting the exact relevance score may not be necessary because the final objective is to produce a ranked list of documents.

In the pairwise category, the ranking model does not try to accurately predict the exact real-valued relevance score of a query-document pair; instead, the objective of the training is to focus on the relative order between two documents for a given query. So, by training using the pairwise category, the ranking model tries to produce a ranked list of documents. In the pairwise approach, ranking is reduced to a binary classification to predict which of two documents is more relevant to a given query. Many pairwise approaches are proposed in the literature including methods that are based on support vector machines (Herbrich et al., 2000; Joachims, 2002), neural networks (Burges et al., 2005), Boosting (Freund et al., 1998), and other machine learning algorithms (Zheng et al., 2007, 2008). For a given query, the number of pairs is quadratic, which means that if there is an imbalance in the relevance judgments where more groundtruth relevance scores are available for a particular query, this imbalance will be magnified by the pairwise approach. In addition, the pairwise method is more sensitive to noise than the pointwise method because a noisy relevance score for a single query-document pair leads to multiple mislabeled document pairs.

The third learning category for ad-hoc retrieval is known as the listwise category, proposed by Cao et al. (2007). In the listwise category, the input to the ranking model is the entire set of documents that are associated with a given query in the training data. Listwise approaches can be divided into two types. In the first, the loss function is directly related to evaluation measures (Chakrabarti et al., 2008; Chapelle & Wu, 2009; Qin et al., 2009). So, they directly optimize for a ranking metric such as NDCG, which is more challenging because these metrics are often not differentiable with respect to the model parameters. Therefore, these metrics are relaxed by approximation to make computation efficient. For the second type, the loss function is differentiable, but it is not directly related to the evaluation measures (Cao et al., 2007; Huang & Frey, 2008; Volkovs & Zemel, 2009). For example, in ListNet (Cao et al., 2007), the probability distribution of permutations is used to define the loss function. Since a ranking list can be seen as a permutation of documents associated with a given query, a model representing the probability distribution of permutations, like the Plackett-Luce (Plackett, 1975) model, can be applied for ranking in ListNet.

## 6 Model categories

We discuss neural ranking models that are proposed in the document retrieval literature based on multiple dimensions. These dimensions capture the neural components and design of the proposed methods in order to better understand the benefits of each design principle.

### 6.1 Representation-focused models vs. interaction-focused models

When extracting features from a query-document pair, the feature extractor $F$ can be applied separately to the query and document, or it can be applied to the interaction between the query and document.

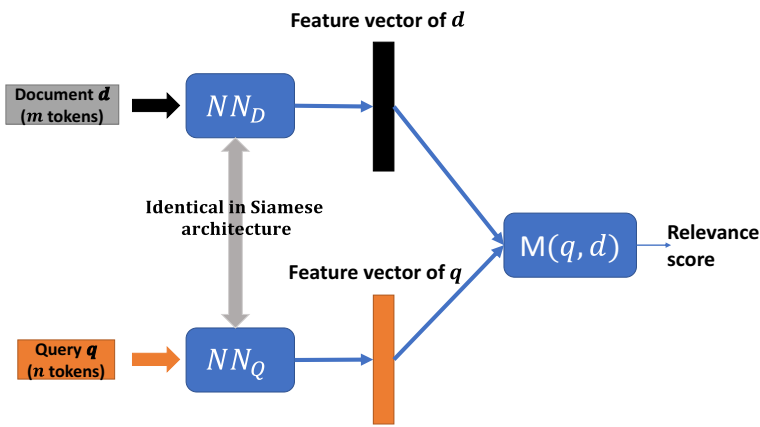

**Fig. 2** Overview of the general architecture of representation-focused models. Two deep neural networks are used to map the query and document to feature vectors. A ranking function $M$ is used to map the feature vectors of the query and document to a real-valued relevance score

### 6.1.1 Representation-focused models

The general framework of representation-focused models is shown in Fig. 2. In representation-focused models, two independent neural network models $NN_Q$ and $NN_D$ map the query $q$ and the document $d$, respectively, into feature vectors $NN_Q(q)$ and $NN_D(d)$. Thus the feature extractor $F$ for a query-document pair is given by:

$$F(q,d) = (NN_Q(q), NN_D(d)) \quad (3)$$

In the particular case where $NN_Q$ and $NN_D$ are identical, the neural architecture is considered to be Siamese (Bromley et al., 1993).

The relevance score of the query-document pair is calculated using a simple $M$ function like cosine similarity, or a Multi-Layer Perceptron (MLP) between the representations of query and document:

$$M(q,d) = cosine(NN_Q(q), NN_D(d)); \quad or \quad M(q,d) = MLP([NN_Q(q); NN_D(d)])$$

The representation-focused model extracts a good feature representation for an input sequence of tokens using deep neural networks. Huang et al. (2013) proposed the first deep neural ranking model for web search using query-title pairs. The proposed model, called Deep Structured Semantic Model (DSSM), is based on the Siamese architecture (Bromley et al., 1993), which is composed of a deep neural network model that extracts features from query and document independently. The deep model is composed of multiple fully connected layers that are used to map high-dimensional textual sparse features into low-dimensional dense features in a semantic space.

In order to capture local context in a given window, Shen et al. (2014b) proposed a Convolutional Deep Structured Semantic Model (C-DSSM) in which a CNN is used instead of feed-forward networks in the Siamese architecture. The feature extractor $F$ is composed of a CNN that is applied to a letter-trigram input representation, then a max-pooling layer is used to form a global feature vector, while $M$ is the cosine similarity function. CNNs have also been used in ARC-I (Hu et al., 2014) to extract feature representations of the query and document. Each layer of ARC-I contains convolutional filters and max-pooling. The input to ARC-I is any pre-trained word embedding. In order to decrease the dimension of representation, and to filter low signals, a max-pooling of size two is applied for each feature map. After applying several layers of CNN filters and max-pooling, ARC-I forms a final feature vector $NN_Q(q)$ and $NN_D(d)$ for query and document, respectively ($NN_Q$ and $NN_D$ are identical because ARC-I follows the Siamese architecture). $NN_Q(q)$ and $NN_D(d)$ are concatenated and fed to a MLP to predict the relevance score. CNN is also the main component in the deep neural networks introduced in Convolutional Neural Tensor Network (Qiu & Huang, 2015) and Convolutional Latent Semantic Model (Shen et al., 2014a).

Recurrent neural networks (RNN), especially the Long Short-Term Memory (LSTM) model (Hochreiter & Schmidhuber, 1997), have been successful in learning to represent each sentence as a fixed-length feature vector. Mueller and Thyagarajan (2016) proposed Manhattan LSTM (MaLSTM) which is composed of two LSTM models as feature extractors. $M$ is a simple similarity measure. LSTM-RNN (Palangi et al., 2016) is also composed of two LSTM, where $M$ is the cosine similarity function. In order to capture richer context, bidirectional LSTM (bi-LSTM) (Schuster & Paliwal, 1997) utilizes both previous and future contexts by processing the sequence data from two directions using two LSTM. Bi-LSTM is used in MV-LSTM (Wan et al., 2016a) to capture the semantic matching in each position of the document and query by generating positional sentence representations. The next step in MV-LSTM is to model the interactions between the generated features using the tensor layer (Socher et al., 2013b, c). The matching between query and document is usually captured by extracting the strongest signals. Therefore, k-max pooling (Kalchbrenner et al., 2014) is used to extract the top *k* strongest interactions in the tensor layer. Then, a MLP is used to calculate the relevance score.

### 6.1.2 Interaction-focused models

Models in the representation-focused group defer the interaction between two inputs until extracting individual features, so that there is a risk of missing important matching signals in the document retrieval task. The interaction-based models start by building local interactions for a query-document pair using simple representations, then train a deep model to extract the important interaction patterns between the query and document. The general framework for interaction-focused models is shown in Fig. 3. The interaction-based models capture matching signals between query and document in an early stage.

In interaction-focused models, *F* captures the interactions between query and document. For example, Guo et al. (2016) introduced a Deep Relevance Matching Model (DRMM) to perform term matching using histogram-based features. The interaction matrix between query and document is computed using pairwise cosine similarities between the embeddings of query tokens and document tokens. DRMM builds a histogram-based feature to extract matching patterns from different levels of interaction signals rather than different positions. In order to control the contribution of each query token to the final relevance score, the authors propose a term gating network with a softmax function.

The histogram feature in DRMM (Guo et al., 2016) is computed based on a hard assignment of cosine similarities between a given query token and the document tokens. This histogram-based feature counts the total number of document tokens with a similarity to the query token that falls within the predefined bin's range of the histogram. The histogram-based representation is not differentiable for the purpose of updating the ranking model parameters in the back-propagation phase, and not computationally efficient. To solve this problem, kernel pooling for soft-match signals is used in K-NRM (Xiong et al., 2017b). Pairwise cosine similarities are compared against a set of *K* kernels, where each kernel represents a normal distribution with a mean and standard deviation. Then, kernel pooling is applied to summarize the cosine similarities into a soft-matching feature vector of dimension $K$; intuitively, this vector represents the probabilities that the similarities came from the distribution specified by each kernel. The final feature vector is computed by summing the soft-matching feature vectors of query tokens.

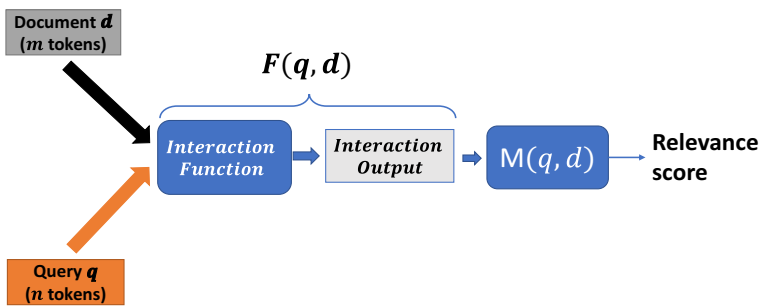

**Fig. 3** Overview of the general architecture of interaction-focused models. An interaction function is used to map the query and document to an interaction output. A ranking function *M* is used to map the interaction output to a real-valued relevance score

Cosine similarity interaction matrix is also used in Hierarchical Neural maTching model (HiNT) (Fan et al., 2018), aNMM (Yang et al., 2016a), MatchPyramid (Pang et al., 2016a, b), and DeepRank (Pang et al., 2017). In addition to cosine similarity, other forms of similarities include dot product and indicator function which are used in HiNT and MatchPyramid, and Gaussian Kernel that is introduced in the study of MatchPyramid (Pang et al., 2016a) using multiple interaction matrices.

Different architectures are used for feature extractor $F$ to build the query-document interactions, and for the ranking model $M$ to extract matching signals from interactions of query and document tokens.

*LSTM-based ranking models*: As in representation-based models, LSTM is used in multiple neural ranking models (Fan et al., 2018; He & Lin, 2016; Jaech et al., 2017). He and Lin (2016) use bi-LSTMs for context modelling of text inputs. So, each input is encoded to hidden states using weight-shared bi-LSTMs. In the ranking model proposed by Jaech et al., (2017), two independent bi-LSTMs without weight sharing are used to map query and document to hidden states. The query and document have different sequential structures and vocabulary which motivates encoding each sequence with independent LSTM. Fan et al., (2018) proposed a variant of the HiNT model that accumulates the signals from each passage in the document sequentially. In order to achieve that, the feature vectors of all passages are fed to an LSTM model to generate a hidden state for each passage. Then, a dimension-wise k-max pooling layer is applied to select top-k signals.

*GRU-based ranking models*: A 2-D GRU is an extension of GRU for two-dimensional data such as an interaction matrix. It scans the input data from top-left to bottom-right (and bottom-right to top-left in case of bidirectional 2D-GRU) recursively. A 2-D GRU is used in Match-SRNN (Wan et al., 2016b) to accumulate the matching signals.

In the neural ranking model that is proposed by Fan et al. (2018), given query-document interaction tensors representing semantic and exact matching signals, a spatial GRU is used to extract the relevance matching evidence. GRU is applied to multiple passages in a document in order to form the matching signal feature vectors. Then, k-max pooling extracts the strongest signals from all passages of a document.

As in Match-SRNN (Wan et al., 2016b) and MatchPyramid (Pang et al., 2016a, b), DeepRank (Pang et al., 2017) has an input interaction tensor between query and document. The input tensor is fed to the GRU network to compute a query-centric feature vector.

*CNN-based ranking models*: A CNN is used in multiple interaction focused models including (Dai et al., 2018; Hui et al., 2017; Jaech et al., 2017; Lan & Xu, 2018; McDonald et al., 2018; Nie et al., 2018; Pang et al., 2016b; Tang & Yang, 2019). Hu et al. (2014) presented ARC-II which is an interaction-based method. ARC-II lets the query and document meet before their feature vectors are fully developed by operating directly on the interaction matrix. Given a sliding window $k_1$ that scans both query and document by taking overlapping sub-sequences of tokens with length $k_1$, a 1−D convolution is applied to all sequences that are formed by concatenating tokens from the sliding window of query and document. The next layers are composed of a 2 × 2 non-overlapping max-pooling and 2−D convolution. Several max-pooling and CNN layers can be added to the model, and the final feature vector is fed to a MLP to predict the query-document relevance score.

Hui et al. (2017) argued that retrieval methods are based on unigram term matches, and they ignore position-dependent information such as proximity and term dependencies. The authors proposed a Position-Aware Convolutional Recurrent Relevance (PACRR) matching model to capture information about the position of a query term and how it interacts with tokens from a document. In order to extract local matching patterns from the cosine similarity interaction matrix, the authors applied CNN filters with multiple kernel sizes. A max-pooling is then applied over the depth channel (number of CNN filters) of the feature maps. This operation assumes that only one matching pattern from all filters is important for a given kernel size representing a query and document n-gram size. The final feature vector is computed using a second k-max pooling over the query dimension in order to keep the strongest signals for each query token.

McDonald et al. (2018) proposed a model called PACRR-DRMM that adapts a PACRR model for the DRMM architecture in order to incorporate contextual information of each query token. A PACRR-based document-aware query token encoding is used instead of the histogram-based feature of DRMM. Then, like in DRMM, each PACRR-based feature is passed through a MLP to independently score each query encoding. Finally, the resulting scores are aggregated using a linear layer. Unlike DRMM, PACRR-DRMM does not include a term gating network that outputs a weight for each query token, because the PACRR-based feature already includes inverse document frequency (IDF) scoring for each query token.

Jaech et al. (2017) designed a neural ranking model called Match-Tensor that explores multiple channel representation for the interaction tensor to capture rich signals when computing query-document relevance scores. The similarity between query and document is computed for each channel. Two bi-LSTMs are used to encode the word embedding-based representation of the query and document into LSTM states. The encoded sequences capture the sequential structure of query and document. A 3-D tensor (query, document, and channel dimensions) is then calculated by point-wise product for each query term representation and each document term representation to produce multiple match channels. A set of convolutional filters is then applied to the 3-D tensor in order to predict the query-document relevance score.

The idea of n-gram soft matching is further investigated by Dai et al. (2018) in the Conv-KNRM model. CNN filters are used to compose n-grams from query and document. This leads to an embedding for each n-gram in the query and document. The n-gram embeddings are then fed to a cross-match layer in order to calculate the cosine similarities between query and document n-grams. Similar to Xiong et al. (2017b), kernel pooling is used to build soft matching feature vectors from the cosine similarity matrices.

*MLP-based ranking models*: The objective of ranking models is to predict a real-valued relevance score for a given query-document pair. So, most of the proposed neural architectures contain a MLP layer that is used to map the final feature vector into a real-valued score. In general, the MLP used in neural ranking architectures is a nonlinear function.

### 6.1.3 Representation+interaction models

Retrieval models can benefit from both representation and interaction deep architectures in a combined model. In DUET (Mitra et al., 2017), an interaction-based network, called the local model, and a representation-based network, called the distributed model, are combined in a single deep learning architecture. Figure 4 shows the overview of the combined representation and interaction models for DUET. The local model takes the interaction matrix of query and document, based on patterns of exact matches of query terms in the document, as input. Then, the interaction matrix is passed through a CNN. The output of the convolutional layer is passed through two fully connected layers, a dropout layer, and a final fully connected layer that produces a final relevance score. The distributed model learns a lower-dimensional feature vector for the query and document. A word embedding-based representation is used to encode each query term and document term. A series of nonlinear transformations is then applied to the embedded input. The matching between query and document representations is calculated using element-wise product. The final score under the DUET architecture is the sum of scores from the local and the distributed networks.

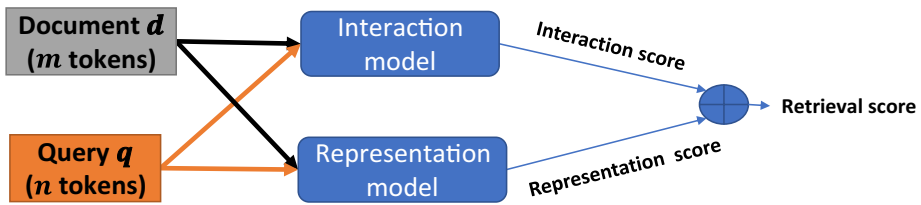

**Fig. 4** Overview of the framework of Representation+Interaction model for DUET. An interaction model and a representation model are used to compute the interaction and representation scores, respectively for a query-document pair. The final score under the DUET is the sum of scores from the interaction and representation models

In recently proposed models, there are more proposed architectures in the interaction-focused category than the representation-focused category. Nie et al. (2018) show in their empirical study that the interaction-based neural architectures generally lead to better results than the representation-focused architectures in information retrieval tasks. Although the representation-focused models offer the advantage of efficient computation by having the same feature vector for a document in all tasks, a static feature representation is unable to capture the matching signals in different tasks and datasets. On the other hand, the interaction-focused neural networks can be computationally expensive, as they require pairwise similarities between embeddings of query and document tokens, but they have the advantage of learning the matching signals from the interaction of two inputs at the very beginning stages.

### 6.2 Context aware-based representation

As suggested by Wu et al. (2007), if there is some relevant information in a document, the relevant information is located around the query terms in the document, which is known as the query-centric assumption. An example of the query-centric assumption is shown in Fig. 5. The query-centric assumption is closely related to human judgement of relevance which consists of three steps (Wu et al., 2007). The first step consists of finding candidate locations in the document for relevant information to the query. Then the objective of the second step is to judge the local relevance of the candidate locations. Finally, the third step consists of aggregating local relevance information to assess the overall relevance of the document to the query.

Inspired by human judgment of relevance, the DeepRank (Pang et al., 2017) architecture considers two information retrieval principles. The first principle is query term importance: a user expresses his request via query terms and some terms are more important than others (Fang et al., 2004). The second principle is the diverse matching requirement of query-centric contexts, which indicates that the distribution of matching

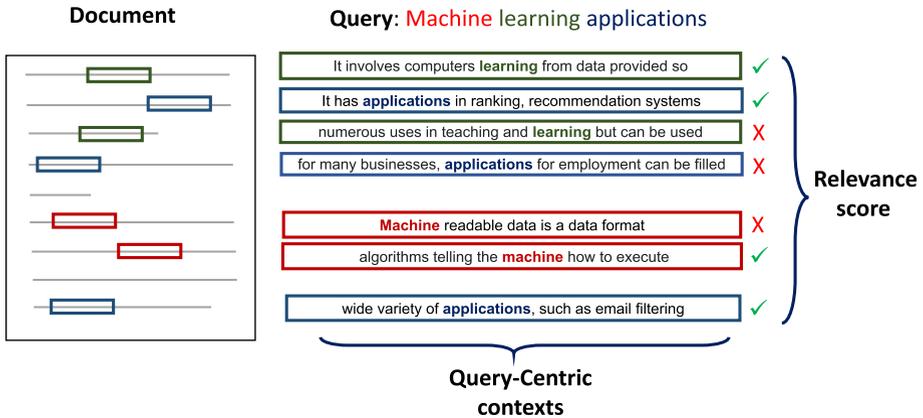

**Fig. 5** Query-centric assumption. The sentences that are used in this example are extracted from Wikipedia. Each query token is shown with a different color that corresponds to the query-centric context in the document. A binary judgement is shown to indicate the relevance between the query and the query-centric context, where red cross denotes not relevant assessment, and the green check mark denotes relevant assessment. The final relevance score is the aggregation of the query-centric relevance scores

signals can be different in a collection of relevant documents. In this context, there are two hypotheses as described by Robertson et al. (1993). The first hypothesis is known as the scope hypothesis which contains long documents with several short and unrelated documents concatenated together. The second hypothesis is known as the verbosity hypothesis, where each document covers a similar topic, but it uses more words to describe the given topic. From the neural architecture perspective, the relevance matching of the verbosity hypothesis category is global because the document contains a single topic. On the contrary, the relevance matching of scope hypothesis can be located at any part of the document, and it only requires a passage of a document to be relevant to a query. A neural architecture should be able to capture local matching patterns. Also, a neural ranking model should incorporate query context into the ranking model in order to solve the ambiguity problem. For example, the query "river bank" should have a lower relevance score with a sentence or passage containing "withdrawal from the bank" despite having the word "bank" in common.

For the first step of the query-centric assumption, DeepRank assumes that the relevance occurs around the exact matching positions of query tokens in a document as shown in Fig. 5. The local relevance signal in the second step is measured using CNN filters with all possible combinations of widths and heights that are applied to the local interaction tensor. DeepRank encodes the position of the query-centric context using the document location that a query token matches exactly; this feature is appended to the CNN feature map to obtain the query-centric feature vector for each exact matching position. Finally, in the third step, DeepRank aggregates the local relevance in two phases. In the first phase, query-centric features of a given query token are passed through LSTM or GRU to obtain a relevance representation for each query token. Then, in the second phase, a Term Gating Network is used to globally aggregate relevance representations of query tokens as in DRMM. DeepRank only considers query context for tokens that have an exact match in a given document, so it ignores query tokens that have similar meaning to a word in a document.

Co-PACRR (Hui et al., 2018) is an alternative to DeepRank that computes the context of each word in a document by averaging the word embeddings of a window around the word, and the meaning of a query by averaging all word embeddings of the query. So, in order to decrease the number of false signals in Co-PACRR due to ambiguity, the extracted matching signals at a given position in a document are adjusted using the similarity between query meaning and context vector of a token in the document. Co-PACRR uses a cascade k-max pooling instead of k-max pooling, which consists of applying k-max pooling at multiple positions of a document in order to capture the information about the locations of matches for documents in the scope hypothesis category. Although the location of matching signals in a document is important, the sequential order of query tokens is ignored in Co-PACRR. It shuffles the feature map with respect to query tokens before feeding the flattened result to dense layers. Shuffling enables the model to avoid learning a signal related to the query term position after extracting n-gram matching patterns. For example, we mentioned in Sect. 4 that the number of tokens in a given query is $n$ which indicates that for models that only support fixed size inputs, short queries are padded to $n$ tokens. Therefore, without shuffling, a model can learn that query tokens at the tail are not important because of padding of short queries, and this leads to ignoring some relevant tokens when calculating relevance score for longer queries.

More coarse-grained context than the sliding window strategy can be used to capture local relevance matching signals in the scope hypothesis category, where a system can divide a long document into passages and collect signals from the passages in order to make a final relevance assessment on the whole document. Callan (1994) discussed how passages should be defined, and how they are incorporated in the ranking process of the entire document. The HiNT (Fan et al., 2018) model is based on matching query and passages of a given document, and then it aggregates the passage-level matching signals using either k-max pooling or bi-LSTM or both.

We can distinguish two types of matches between query and document tokens. The first type is the context-free match where the similarity between a given document token and query token is computed regardless of the context of the tokens. Examples of the context-free match are the cosine similarity between a query token embedding and a document token embedding, and exact matching. The second type is the context-sensitive match where the context of a given document token is matched against the context of a query token. In order to capture contextual information of a given token in document and query when calculating the matching similarity, a context-sensitive word embedding can be used to encode tokens. When a query has many context-sensitive matches with a given document, it is likely that the document is relevant to the query. The idea of context-sensitive matching is incorporated into the neural ranking model which is proposed by McDonald et al. (2018) to extend the DRMM model by using bi-LSTM to obtain context-sensitive embedding, and to capture signals of high density context-sensitive matches.

Matching the contexts can be in two directions: matching the overall context of a query against each token of a document, and matching the overall context of a document against each token of a query. To match the contextual representation of the query and document in the two directions, a neural ranking model defines a matching function to compute the similarity between contexts. Wang et al. (2017b) proposed a Bilateral Multi-Perspective Matching (BiMPM) model to match contexts of two sentences. After encoding each token in a given sentence using the bi-LSTM network, four matching strategies are used to compare the contextual information. The proposed matching strategies differ on how to aggregate the contextual information of the first input, which is matched against each time step of the second input (and vice versa). The four matching strategies generate eight vectors in each time step (there is forward and backward contextual information) which are concatenated and fed to a second bi-LSTM model to extract two feature vectors from the last time step of forward and backward LSTM. The whole process is repeated to match contexts in the inverse direction and extract two additional feature vectors. The four final feature vectors are concatenated and fed to a fully connected network to predict a real-valued score.

### 6.3 Attention-based representation

McDonald et al. (2018) proposed a model, called Attention-Based ELement-wise DRMM (ABEL-DRMM), that takes advantage of the context-sensitive embedding and attention weights. Any similarity measure between the term encodings of the query and document tokens already captures contextual information due to the context-sensitive embedding. In ABEL-DRMM, the first step is to calculate attention weights for each query token against document tokens using softmax of cosine similarities. Then, the attention-based representation of a document is calculated using the attention weights and the embedding of document tokens. The document-aware query token encoding is then computed using element-wise multiplication between the query token embedding and the attention-based representation of a document. Finally, in order to compute the relevance score, the document-aware query token encodings of all query tokens are fed to the DRMM model.

A symmetric attention mechanism or co-attention can focus on the set of important positions in both textual inputs. Kim et al. (2019) incorporate the attention mechanism in Densely-connected Recurrent and Co-attentive neural Network (DRCN). DRCN uses residual connections (He et al., 2016), like in Densenet (Huang et al., 2017), in order to build higher level feature vectors without exploding or vanishing gradient problems. When building feature vectors, the concatenation operator is used to preserve features from previous layers for final prediction. The co-attentive network uses the attention mechanism to focus on the relevant tokens of each text input in each RNN layer. Then, the co-attentive feature vectors are concatenated with the RNN feature vectors of every token in order to form the DRCN representation.

The idea of concatenating feature vectors from previous layers before calculating attention weights is also explored by Zhang et al. (2019) in their proposed model DRr-Net. DRr-Net includes an attention stack-GRU unit to compute an attention-based representation for both inputs that capture the most relevant parts. It also has a Dynamic Re-read (DRr) unit that can focus on the most important word at each step, taking into consideration the learned information. The selection of important words in the DRr unit is also based on attention weights.

Using multiple attention functions in matching with a word-by-word attention (Rocktäschel et al., 2016) can better capture the interactions between the two inputs. Multiple attention functions are proposed in the literature: e.g., bilinear attention function (Chen et al., 2016) and concatenated attention function (Rocktäschel et al., 2016). The bilinear attention is based on computing the attention score between the representations of two tokens using the bilinear function. The concatenated attention starts by summing the two words' representations, and then uses vector multiplication to compute attention weight. Tan et al. (2018) propose a Multiway Attention Network (MwAN) using multiple attention functions for semantic matching. In addition to the bilinear and concatenated attention functions, MwAN uses two other attention functions which are the element-wise dot product and difference of two vectors. The matching signals from multiple attention functions are aggregated using a bi-directional GRU network and a second concatenated attention mechanism to combine the four representations. The prediction layer is composed of two attention layers to output the final feature vector that is fed to a MLP in order to obtain the final relevance score.

The use of multiple functions in attention is not limited to the calculation of attention weights. Multiple functions can be used to compare the embedding of a given token with its context generated using the attention mechanism. Wang and Jiang (2017) use several comparison methods in order to match the embedding of a token and its context. These comparison methods include neural network layer, neural tensor network (Socher et al., 2013d), Euclidean distance, cosine similarity, and element-wise operations for vectors.

In addition to being used in LSTM models, the attention mechanism has been beneficial to CNN models. Yin et al. (2015) proposed an Attention Based Convolutional Neural Network (ABCNN) that incorporates the attention mechanism on both the input layer and the feature maps obtained from CNN filters. ABCNN computes attention weights on the input embedding in order to improve the feature map computed by CNN filters. Then, ABCNN computes attention weights on the output of CNN filters in order to reweight feature maps for the attention-based average pooling.

### 6.4 External knowledge and feedback

Many methods have been developed to incorporate knowledge bases into retrieval components. For example, the description of entities can be used to have better term expansion (Xu et al., 2009), or to expand queries to have better ranking features (Dalton et al., 2014). Queries and documents are connected through entities in the knowledge base to build a probabilistic model for document ranking based on the similarity to entity descriptions (Liu & Fang, 2015). Other researchers have extended bag-of-word language models to include entities (Raviv et al., 2016; Xiong et al., 2016).

Knowledge bases can be incorporated in neural ranking models in multiple ways. The AttR-Duet system (Xiong et al., 2017a) uses the knowledge base to compute an additional entity representation for document and query tokens. There are four possible interactions between word-based feature vectors and entity-based feature vectors based on the inter- and intra-space matching signals. The textual attributes of entities such as the description and names are used in inter-space interactions of entities and words. AttR-Duet learns entity embeddings from the knowledge graph (Wang et al., 2017) and uses a similarity measure on these embeddings to determine intra-space interactions of entities. The ranking model combines the four types of interactions with attention weights to reduce the effect of entity linking mistakes on the ranking score.

Knowledge graph semantics can be incorporated into interaction-based neural ranking models as proposed by Liu et al., (2018). The authors propose combining three representations for an entity in a knowledge graph: entity embedding, description embedding, and type embedding. The final entity representation is integrated into kernel pooling-based ranking models, such as K-NRM (Xiong et al., 2017b) and Conv-KNRM (Dai et al., 2018), with word-level interaction matrices to compute the final relevance score of a query-document pair. Shen et al., (2018) develop a knowledge-aware attentive neural-ranking model which learns both context-based and knowledge-based sentence representations. The proposed method leverages external knowledge from the KB using an entity mention step for tokens in the inputs. A bi-LSTM computes a context-aware embedding matrix, which is then used to compute context-guided knowledge vectors. These knowledge vectors are multiplied by the attention weights to compute the final context-guided embedding for each token.

Inspired by traditional ranking techniques that use Pseudo Relevance Feedback (PRF) (Hedin et al., 2009) to improve retrieval results, PRF can be integrated in neural ranking models as in the model proposed by Li et al. (2018). The authors introduce a Neural framework for Pseudo Relevance Feedback (NPRF) that uses two ranking models. The first model computes a query-based relevance score between the query and document corpus and extracts the top candidate documents for a given query. Then, the second model computes a document-based relevance score between the target document and the candidate documents. Finally, the query-based and document-based relevance scores are combined to compute the final relevance score for a query-document pair.

### 6.5 Deep contextualized language model-based representations

The sentence pair classification setting is used to solve the document retrieval task. The overview of BERT for the document retrieval is shown in Fig. 6. In general, the input sequence to BERT is composed of the query $q$ and selected tokens $s_d$ from the document $d$: [[CLS], $q$, [SEP], $s_d$, [SEP]]. The selected tokens can be the whole document, sentences, passages, or individual tokens. The hidden state of the [CLS] token is used for the final retrieval score prediction.

While BERT has been successfully applied to Question-answering (QA), applying BERT to ad-hoc retrieval of documents comes with the challenge of having significantly

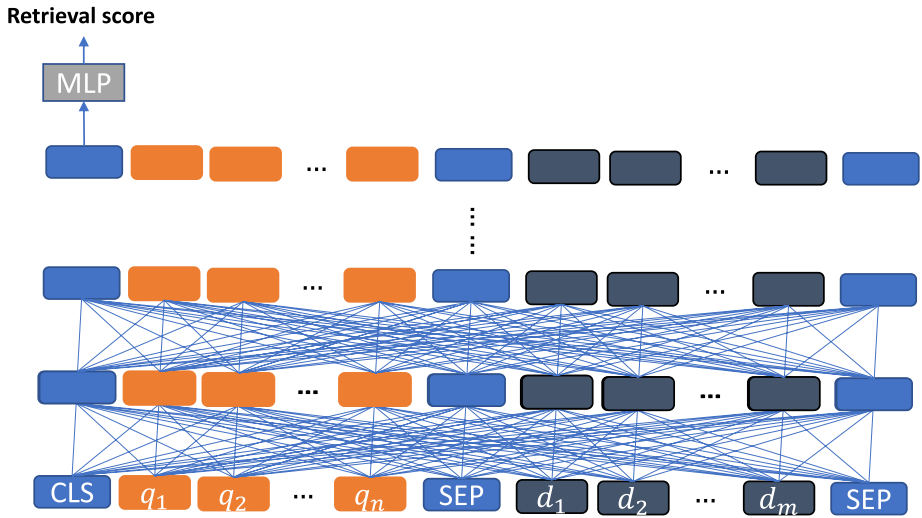

**Fig. 6** Overview of BERT model for document retrieval. The input sequence to BERT is composed of the query $q = q_1 q_2 \ldots q_n$ (shown with orange color in the first layer) and selected tokens $s_d = d_1 d_2 \ldots d_m$ (shown with dark blue color in the first layer) from the document $d$. The BERT-based model is formed of multiple layers of Transformer blocks where each token attends to all tokens in the input sequence for all layers. From the second to the last layer, each cell represents the hidden state of the corresponding token which is obtained from the Transformer. The query and document tokens are concatenated using the [SEP] token, and [CLS] is added to the beginning of the concatenated sequence. The hidden state of [CLS] token is used as input to MLP in order to predict the relevance score (Color figure online)

longer documents than BERT allows (BERT cannot take input sequences longer than 512 tokens). Yang et al. (2019a) proposed to address the length limit challenge by dividing documents into sentences and applying BERT to each of these sentences. The sentence-level representation of a document is motivated by recent work (Zhang et al., 2018) which shows that a single excerpt of a document is better than a full document for high recall in retrieval. In addition, using sentence-level representation is related to research in passage-level document ranking (Liu & Croft, 2002). For each document, its relevance to the query can be predicted using the maximum relevance of its component sentences which is denoted as the best sentence. Yang et al. (2019a) generalize the best sentence concept by choosing the top-$k$ sentences from each document based on the retrieval score calculated by BERT for sentence pair classification setting. A weighted sum of the top-$k$ sentence-level scores, which are computed by BERT, is then applied to predict the retrieval score of the query-document pair. In the training phase, BERT is fine-tuned on microblog data or QA data, and the results show that training on microblog is more effective than QA data for ad-hoc document retrieval (Yang et al., 2019a). The hidden state of the [CLS] token is also used by Nogueira and Cho (2019) to rank candidate passages.

For long document tasks such as document retrieval on ClueWeb09-B (Dai et al., 2018), XLNet (Yang et al., 2019b) uses TransformerXL (Dai et al., 2019) instead of BERT. TransformerXL uses a relative positional encoding and segment recurrence mechanism to capture longer-term dependency. XLNet (Yang et al., 2019b) results in a performance gain around 1.87% for NDCG@20 compared to the BERT-based model.

Qiao et al. (2019) explore multiple ways to fine-tune BERT on two retrieval tasks: TREC Web Track ad-hoc document ranking and MS MARCO (Nguyen et al., 2016) passage re-ranking. Four BERT-based ranking models are proposed which are related to both representation and interaction based models using the [CLS] embedding, and also the embeddings of each token in the query and document. The authors show that BERT works better with pairs of texts that are semantically close. However, as mentioned before, queries and documents can be very different, especially in their lengths and can benefit from relevance matching techniques.

BERT contains multiple Transformer layers, where the deeper the layer, the more contextual information is captured. BERT can be used in the embedding layer to extract tensor contextualized embeddings for both query and document. Then, the interactions between query and document is captured by computing an interaction tensor from the embedding tensors of both the query and document. Finally, a ranking function maps the interaction tensor to a real-valued relevance score. The overview of the BERT model that is used as an embedding layer is shown in Fig. 7.

In order to capture both relevance and semantic matching, MacAvaney et al. (2019) propose a joint model that incorporates the representation of [CLS] from the query-document pair into existing neural ranking models (DRMM (Guo et al., 2016), PACCR (Hui et al.,

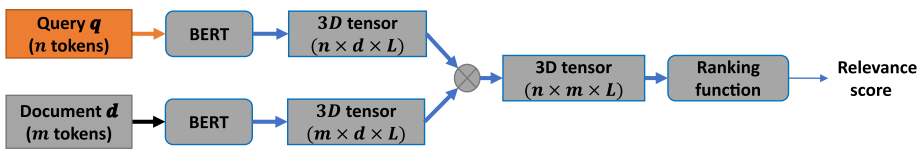

**Fig. 7** Overview of BERT model used as an embedding layer for document retrieval. $L$ is the number of Transformer layers in BERT. Pairwise cosine similarities are computed per layer to obtain a $3D$ interaction tensor

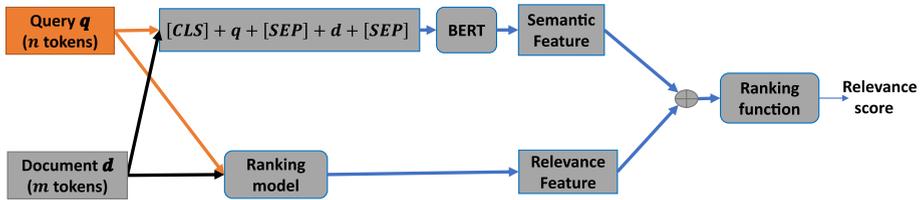

**Fig. 8** Overview of a possible joint model for document retrieval that incorporates both the semantic and relevance matching. BERT is used as a semantic matching component, where the embedding of the [CLS] token is considered as a semantic feature. An existing relevance-based neural ranking model extracts the relevance feature from a query-document pair

2017), and K-NRM (Xiong et al., 2017b)). The overview of a simple example of a joint model is shown in Fig. 8. The representation of the [CLS] token provides a strong semantic matching signal given that BERT is pretrained on the next-sentence prediction. As we explained previously, some of the neural ranking models, like DRMM, capture relevance matching for each query term based on the similarities with the document tokens. For ranking models, MacAvaney et al. (2019) use pretrained contextual language representations as input, instead of the conventional pretrained word vectors to produce a context-aware representation for each token from the query and document.

Dai and Callan (2019) augment the BERT-based ranking model with the search knowledge obtained from search logs. The authors show that BERT benefits from tuning on the rich search knowledge, in addition to the language understanding knowledge which is obtained from training BERT on query-document pairs.

Nogueira et al. (2019) propose a multi-stage ranking architecture. The first stage consists of extracting the candidate documents using BM25. In this stage, recall is more important than precision to cover all possible relevant documents. The irrelevant documents can be discarded in the next stages. The second stage, called monoBERT, uses a pointwise ranking strategy to filter the candidate documents from the first stage. The classification setting of BERT with sentence pairs is used to compute the relevance scores. The third stage, called duoBERT, is a pairwise learning strategy that computes the probability of a given document being more relevant than another candidate document. Documents from the second stage are ranked using duoBERT relevance scores in order to obtain the final ranked list of documents. The input to duoBERT is the concatenation of query, first document, and second document, where [SEP] is added between the sentences, and [CLS] is added to the beginning of the concatenated sentence.

As explained earlier, exact matching is an important matching signal in traditional IR models, and relevance matching-based neural ranking models incorporate the exact matching signal to improve retrieval results. In order to directly incorporate exact matching signal in the sentence pair classification setting of BERT for document retrieval, Boualili et al. (2020) proposed to mark the start and end of exact matching query tokens in a document with special markers.

## 7 Neural ranking models features

To summarize the neural models from the five categories, we propose nine features that are frequently presented in the neural ranking models.

1. *Symmetric*: We describe a neural ranking architecture as symmetric if the relevance score does not change if we change the order of inputs (query-document or document-query pairs). In other words, for a given query and document, there are no special computations that are applied only to the query or document.
2. *Attention*: This dimension characterizes neural ranking models that have any type of attention mechanisms.
3. *Ordered tokens*: The sequential order of tokens is preserved for both query and document when computing the interaction tensors between query and document, and the final feature vector representing the query-document pair.
4. *Representation*: This feature characterizes neural ranking models that extract features from the query and document separately, and defer the interactions between the features to the ranking function.

    – *Without weights sharing (W)*: Two independent deep neural networks without weights sharing are used to extract features from queries and documents.
    – *With weights sharing or Siamese (S)*: The neural ranking model has Siamese architecture as described by Bromley et al., (1993), where the same deep neural network is used to extract features from both query and document.

5. *Interaction*: This feature characterizes neural ranking models that build local interactions between query and document in an early stage, so it can be considered as the mutually exclusive category of representation models.
6. *Injection of contextual information*: Depending on when to inject contextual information, we can distinguish two cases.

    – *Early injection of contextual information (E)*: Some neural ranking models incorporate contextual information in the embedding phase by considering context-sensitive embeddings (for example the embedding that is computed using LSTM).
    – *Late injection of contextual information (L)*: Some neural ranking models defer injecting the contextual information until computing the interaction tensors (for example, applying n-gram convolutions on the interaction tensor to incorporate contextual information).

7. *Exact matching*: This feature means that the neural ranking model includes the exact matching signal when calculating relevance score.
8. *Incorporate external knowledge bases (KB)*: This feature characterizes neural ranking models that incorporate external knowledge bases to predict query-document relevance score.
9. *Deep language models (LM)*: This feature refers to the use of deep contextualized language models to compute query-document relevance scores. We can distinguish two cases for deep LM.

    – *Deep LM in embedding layer (Em)*: Deep contextualized language models are used as a context-sensitive embedding to compute a word embedding tensor (because Deep LM have multiple layers) for both query and document. Such models necessarily have the *early injection of contextual information* property.
    – *Deep LM as a semantic matching component (Se)*: Deep contextualized language models are used as a semantic matching component in a neural ranking model. For example, BERT is pretrained on the next sentence prediction so that it captures semantic matching signals. The same for ELMo which is pretrained on the next token prediction.

**Table 1** Overview of Neural Ranking Models

| Method | symmetric | attention | ordered tokens | representation | interaction | injection of CI | exact matching | KB | Deep LM |
|---|---|---|---|---|---|---|---|---|---|
| DSSM (Huang et al., 2013) | ✓ | | | S | | | | | |
| C-DSSM (Shen et al., 2014b) | ✓ | | ✓ | S | | E | | | |
| ARC-I (Hu et al., 2014) | ✓ | | ✓ | S | | E | | | |
| ARC-II (Hu et al., 2014) | ✓ | | ✓ | | ✓ | L | | | |
| ABCNN-2 (Yin et al., 2015) | ✓ | ✓ | ✓ | S | | E | | | |
| MV-LSTM (Wan et al., 2016a) | ✓ | | ✓ | | ✓ | E | | | |
| MaLSTM (Mueller and Thyagarajan 2016) | ✓ | | ✓ | S | | E | | | |
| DRMM (Guo et al., 2016) | | | | | ✓ | | ✓ | | |
| Hybrid of ConvNet and bi-LSTM (He and Lin 2016) | ✓ | | ✓ | | ✓ | E | | | |
| MatchPyramid (Pang et al., 2016b) | ✓ | | ✓ | | ✓ | L | | | |
| DUET (Mitra et al., 2017) | | | | W | ✓ | L | ✓ | | |
| Compare-Aggregate network (Wang and Jiang 2017) | | ✓ | ✓ | | ✓ | E | | | |
| K-NRM (Xiong et al., 2017b) | | | | | ✓ | | ✓ | | |
| Match-Tensor (Jaech et al., 2017) | ✓ | ✓ | ✓ | | ✓ | E | ✓ | | |
| AttR-Duet (Xiong et al., 2017a) | | | | | ✓ | | | ✓ | |
| PACRR (Hui et al., 2017) | | | | | ✓ | L | ✓ | | |
| DeepRank (Pang et al., 2017) | | | ✓ | | ✓ | L | ✓ | | |
| BiMPM (Wang et al., 2017b) | ✓ | ✓ | ✓ | | ✓ | E | | | |
| Co-PACRR (Hui et al., 2018) | | | | | ✓ | E | ✓ | | |
| Inter-2D-2L (Nie et al., 2018) | ✓ | | ✓ | | ✓ | L | | | |
| Conv-KNRM (Dai et al., 2018) | | | | | ✓ | E | ✓ | | |
| MwAN (Tan et al., 2018) | | ✓ | ✓ | | ✓ | E | | | |
| POSIT-DRMM (McDonald et al., 2018) | | | | | ✓ | E | ✓ | | |
| HiNT (Fan et al., 2018) | | | ✓ | | ✓ | L | | | |
| PACRR-DRMM (McDonald et al., 2018) | | | | | ✓ | L | ✓ | | |
| ABEL-DRMM (McDonald et al., 2018) | | ✓ | | | ✓ | | ✓ | | |
| EDRM (Liu et al., 2018) | | ✓ | ✓ | | ✓ | E | | ✓ | |

**Table 1** (continued)

| Method | symmetric | attention | ordered tokens | representation | interaction | injection of CI | exact matching | KB | Deep LM |
|---|---|---|---|---|---|---|---|---|---|
| KABLSTM (Shen et al., 2018) | ✓ | ✓ | ✓ | | ✓ | E | | ✓ | |
| NPRF (Li et al., 2018) | | | | | ✓ | | ✓ | | |
| DeepTileBars (Tang and Yang 2019) | | | ✓ | | ✓ | L | | | |
| DRCN (Kim et al., 2019) | ✓ | ✓ | ✓ | | ✓ | E | | | |
| DRr-Net (Zhang et al., 2019) | ✓ | ✓ | ✓ | | ✓ | E | | | |
| BERT (sentence pair classification) (Yang et al., 2019a) | | ✓ | ✓ | | ✓ | E | | | Se |
| BERT as a passage re-ranker (Nogueira and Cho 2019) | | ✓ | ✓ | | ✓ | E | | | Se |
| BERT (Term-Trans) (Qiao et al., 2019) | ✓ | ✓ | ✓ | | ✓ | E | | | Em |
| Joint BERT (MacAvaney et al., 2019) | | ✓ | ✓ | | ✓ | E | ✓ | | Se |
| Contextualized similarity tensors (MacAvaney et al., 2019) | ✓ | ✓ | ✓ | | ✓ | E | | | Em |
| Augmented BERT (Dai and Callan 2019) | | ✓ | ✓ | | ✓ | E | | | Se |
| monoBERT+duoBERT (Nogueira et al., 2019) | | ✓ | ✓ | | ✓ | E | | | Se |
| TKL (Hofstätter et al., 2020) | | ✓ | | | ✓ | E | ✓ | | |
| MarkedBERT (Boualili et al., 2020) | | ✓ | ✓ | | ✓ | E | ✓ | | Se |
| BERT-based ranking model (Zhan et al., 2020a) | | ✓ | ✓ | | ✓ | E | | | Se |
| ColBERT (Khattab and Zaharia 2020) | | ✓ | ✓ | | ✓ | E | | | Em |

Table 1 shows the main IR features of each neural ranking model from all proposed categories. The neural ranking models are sorted in chronological order based on the publication year. Unsurprisingly, in recent years, there have been more proposed neural ranking models that are based on BERT because deep contextualized language models achieve state-of-the-art results in multiple tasks for NLP and IR. Later in the discussion part, we will discuss more research directions to reduce the time and memory complexity of BERT-based ranking models. Except for DUET (Mitra et al., 2017), all the neural ranking models have either the interaction or representation feature.

Recent proposed methods are interaction-based ranking models that prefer building the interactions between query and document in an early stage to capture matching signals. As mentioned in Sect. 6, DUET combines a representation-based model without weights sharing, with an interaction-based model to predict the final query-document relevance score.

Given the recent advances in the embedding layer for deep learning models, early injection of contextual information is a common design choice for multiple neural ranking models. The contextual information is incorporated from the first stage which consists of an embedding layer either by using traditional neural recurrent components such as LSTM or more advanced deep contextualized representations, such as the Transformer.

In general, the models that are proposed primarily for text matching tasks are symmetric because the inputs are homogeneous. On the other hand, many models that are proposed primarily for document retrieval are not symmetric because there are special computations that are applied only to the query or the document. For example, kernel pooling is used in Conv-KNRM (Dai et al., 2018) to summarize the similarities between a given query token and all document tokens. So, this can be seen as a query-level operation that breaks the symmetric property. An example of a document level operation that leads to an asymmetric architecture is included in ABEL-DRMM (McDonald et al., 2018). The attention weights are only computed for each document token against a given query token to produce the attention-based representation of the document. So, the attention mechanism is only applied at the document level, and therefore the neural ranking model is asymmetric. The asymmetric property of some BERT-based ranking models comes from multiple facts. First, BERT is applied to the sentence or passage level of a long document (Yang et al., 2019a; Nogueira & Cho, 2019; Dai & Callan, 2019; Zhan et al., 2020a), so that there are some preprocessing steps that are applied only to the document. Second, some BERT-based models, such as MacAvaney et al. (2019), are combined with existing relevance-based ranking models that are asymmetric, and others, such as Nogueira et al. (2019), include a component for pairwise comparison of documents, so that the joint model is asymmetric in both cases. Third, Boualili et al. (2020) include the exact matching of query tokens into the ranking model which leads to an overall asymmetric architecture. In the case of short documents where BERT can accept the full document and only the BERT-based model is used for ranking, the ranking model is symmetric.

## 8 Beyond document retrieval

The idea of using neural ranking models to rank documents given a user's query can generalize to other retrieval tasks, with different objects to query and rank. In this section, we discuss the analogy between other forms of retrieval tasks and document retrieval. In particular, we describe four retrieval tasks which are: structured document retrieval, Question-Answering, image retrieval, and Ad-hoc video search.

### 8.1 Structured document retrieval

The information retrieval field has presented multiple methods to incorporate the internal organization of a given document into indexing and retrieval steps. The progress in document design and storage has resulted in new representations for documents, known as structured documents (Chiaramella, 2000), such as HTML and XML, where the document has multiple fields. Considering the structure of a document when designing retrieval models can usually improve retrieval results (Wilkinson, 1994). Zamani et al, (2018) proposed a neural ranking model that extracts the document representation from the aggregation of field-level representations and then uses a matching network to predict the final relevance score.

Beyond Web pages, a table itself can also be considered as a structured document. Zhang & Balog, (2018) propose a semantic matching method for table retrieval where various embedding features are used. Chen et al. (2020a) first learn the embedding representations of table headers and generate new headers with embedding features and curated features (Chen et al., 2018) for data tables. They show that the generated headers can be combined with the original fields of the table in order to accurately predict the relevance score of a query-table pair, and improve ranking performance. Trabelsi et al. (2019) proposed a new word embedding of the tokens of table attributes, called MCON, using the contextual information of every table. Different formulations for contexts are proposed to create the embeddings of attribute tokens. The authors argued that the different types of contexts should not all be treated uniformly and showed that data values are useful in creating a meaningful semantic representation of the attribute. In addition to computing word embeddings, the model can predict additional contexts of each table and use the predicted contexts in a mixed ranking model to compute the query-table relevance score. Using multiple and differentiated contexts leads to more useful attribute embeddings for the table retrieval task.

Shraga et al. (2020) use different neural networks to learn different unimodal representations of a table which are combined into a multimodal representation. The final table-query relevance is estimated based on the query representation and multimodal representation. Chen et al. (2020b) first select the most salient items of a table to construct the BERT representation for the table search, where different types of table items and salient signals are tested. The proposed content selection technique improves the performance of ad-hoc table retrieval. On the other hand, using the queries to select the table content can lead to a significant increase in the processing time because extracting data table representations cannot be performed offline. Trabelsi et al. (2020b) proposed to include summary vectors about the contents of the table, both in terms of values in each column and values in selected rows. The summary vectors compress each row and each column into a fixed length feature vector using word embedding of data values.

Inspired by recent progress of transfer learning on graph neural networks, Trabelsi et al. (2020c) proposed to represent a large collection of data tables using graphs. In particular, a knowledge graph representation using fact triples < *subject, predicate, object*> indicates the relations between entities, then R-GCN (Schlichtkrull et al., 2018), which is an extension of GCN for knowledge graphs, is applied on knowledge graphs to learn representations for graph nodes and relations.

### 8.2 Question-answering

Question-answering (QA) (Diefenbach et al., 2018; Lai et al., 2018; Wu et al., 2019) is the task that focuses on retrieving texts that answer a given user's question. The extracted answers can have different lengths, and vary from short text, passage, paragraph or document. QA also includes choosing between multiple choices, and synthesizing answers from multiple resources in case the question looks for multiple pieces of information.

The QA problem presents multiple challenges. The question is expressed in natural language, and the objective is to search for short answers in a document. So only the parts that are relevant to the question should be extracted. The second challenge is the different vocabulary used in questions and answers. So, a QA model needs to capture matching signals based on the intents of the question in order to extract an accurate answer. The third challenge is to gather responses from multiple sources to compose an answer. As in document retrieval, multiple neural ranking models are proposed to retrieve answers that are relevant to a given user's question (Guo et al., 2019; Abbasiyantaeb & Momtazi, 2020; Huang et al., 2020). The neural ranking models for QA cover all five proposed categories for document retrieval with a focus on the semantic matching signal between questions and answers.

### 8.3 Image retrieval

Image retrieval (Zhou et al., 2017) is the task of retrieving images that are relevant to a user's query. Image retrieval has been studied from two directions: text-based and content-based methods. Text-based image retrieval uses annotations, such as the metadata, descriptions and keywords, that are manually added to the image to retrieve images that are relevant to a keyword-based query. The objective of the annotations is to describe the content of the image so that a large collection of images can be organized and indexed for retrieval. Text-based image retrieval is treated as a text-based information retrieval. With the large increase of image datasets and repositories, describing each image content with textual features becomes more difficult, which has led to low precision for text-based image retrieval. In general, it is hard to accurately describe an image using only a few keywords, and it is common to have inconsistencies between image annotations and a user's query.

In order to overcome the limitations of text-based methods, content-based image retrieval (CBIR) (Dharani & Aroquiaraj, 2013; Wan et al., 2015; Zhou et al., 2017) methods retrieve relevant images based on the actual content of the image. In other words, CBIR consists of retrieving similar images to the user's query image. This is known as the query-by-example setting. An example of a search engine for CBIR is the reverse image search introduced by Google. As in representation-focused models for document retrieval, many CBIR neural ranking models (Wiggers et al., 2019; Chung & Weng, 2017) use a deep neural network as a feature extractor to map both the query image and a given candidate from the image collection into fixed-length representations. So, these neural ranking models have the Siamese feature. Then, a simple ranking function, such as cosine similarity, is used to predict the relevance score of a query-image pair. The Siamese architecture is used in multiple CBIR domains such as retrieving aerial images from satellites (Khokhlova et al., 2020) and content-based medical image retrieval (CBMIR) (Chung & Weng, 2017). CBMIR helps clinicians in the diagnosis by exploring similar cases in medical databases. Retrieving similar images for diagnosis requires extracting content-based feature vectors from medical images, such as MRI data, and then identifying the most similar images to a given query image by comparing the extracted features using similarity metrics.

### 8.4 Ad-hoc video search

Ad-hoc video search (Awad et al., 2016) consists of retrieving video frames from a large collection of videos where the retrieved videos are relevant to the user's query. Similar to document retrieval, text-based video retrieval using the filename, text surrounding the video, etc, has achieved high performance for the video retrieval with a simple query that has few keywords (Snoek & Worring, 2009). Recently, researchers have focused on scenarios where the query is more complex and defined as natural language text. In this case, a cross-modal semantic matching between the textual query and the video is captured to retrieve a set of relevant videos.

Two categories are defined for existing methods on complex query-based video retrieval: concept-based (Snoek & Worring 2009; Yuan et al., 2011; Nguyen et al., 2017) and embedding-based (Li et al., 2019; Cao et al., 2019; Miech et al., 2018, 2019) categories. In concept-based methods, visual concepts are used to describe the content of a video. Then, the user's query is mapped to related visual concepts which are used to retrieve a set of videos by aggregating matching signals from the visual concepts. This approach works well for queries where related visual concepts are accurately extracted. However, capturing semantic similarity between videos and long queries by aggregating visual concepts is not accurate because these queries contain complex semantic meaning. In addition, extracting visual concepts for a video and query is done independently. Embedding-based methods propose to map queries and videos into a common space, where the similarity between the embeddings is computed using distance functions, such as cosine similarity.

As in document retrieval, many video retrieval models are representation-focused models where the only difference is the cross-modal characteristic of the neural ranking model: one deep neural network for video embedding and another deep neural network for text embedding. For example, Yang et al. (2020) propose a text-video joint embedding learning for complex-query video retrieval. The text-based network which is used to embed the query has a context-sensitive embedding with LSTM for an early injection of contextual information as in the document retrieval. Consecutive frames in a video have the temporal dependence feature. So, as in textual input, LSTM can be used to capture the contextual information of frames after extracting frame-based features using pretrained CNN. As in BERT where the self-attention is used to capture token interactions, Yang et al. (2020) introduce a multi-head self-attention for video frames. So, this neural ranking model for video retrieval covers multiple proposed categories for the document retrieval which are: representation-focused models, context-aware based representation, and attention-based representation (with both attention and self-attention).

## 9 Lessons learned and future directions

In this section, we summarize the important signals and neural components that are incorporated into the neural ranking models, and we discuss potential research ideas for document retrieval.

### 9.1 What are the important matching signals in document retrieval?

The neural ranking models that are previously described present two important matching techniques: semantic matching and relevance matching (Guo et al., 2016). Semantic matching is introduced in multiple text matching tasks, such as natural language inference, and paraphrase identification. Semantic matching, which aims to model the semantic similarity between the query and the document, assumes that the input texts are homogeneous. Semantic matching captures composition and grammar information to match two input texts which are compared in their entirety. In information retrieval, the QA task is a good scenario for semantic matching, where semantic and syntactic features are important to compute the relevance score. On the other hand, semantic matching is not enough for document retrieval, because a typical scenario is to have a query that contains keywords. In such cases, the relevance matching is needed to achieve better retrieval results.

Relevance matching is introduced by Guo et al., (2016) to solve the case of heterogeneous query and document in ad-hoc document retrieval. The query can be expressed by keywords, so a semantic signal is less informative in this case because the composition and grammar of a keyword-based query are not well defined. In addition, the position of a given token in a query has less importance than the strength of the similarity signal, so some neural ranking models, like DRMM (Guo et al., 2016), do not preserve the position information when computing the query-document feature vector. An important signal in the relevance matching is the exact matching of query and document tokens. In traditional retrieval models, like BM25, exact matching is primarily used to rank a set of documents, and the model works reasonably well as an initial ranker. Incorporating exact matching into neural ranking models can improve the retrieval performance mainly in terms of recall for keyword-based queries because as in traditional ad-hoc document retrieval, the document has more content than the query and the presence of query keywords in a document is an initial indicator of relevance.

From the review of many neural ranking models, we can conclude that both semantic and relevance matching signals are important to cover multiple scenarios of ad-hoc retrieval tasks. This is empirically justified by achieving significant improvements in retrieval results when using neural ranking models that guarantee both matching signals. For example, the joint model, proposed by MacAvaney et al. (2019), combines the representation of [CLS] from BERT and existing relevance-based neural ranking models. This model has a semantic matching signal from [CLS] because BERT is pretrained on the next sentence prediction, and a relevance matching signal from existing neural ranking models. Both the semantic and relevance components are interaction-based neural architectures that generally lead to better results than the representation-based architectures in information retrieval tasks (Nie et al., 2018). The disadvantage of using the BERT model as a semantic matching component is the length limit of BERT which causes difficulties in both training and inference. In general, the length of a document exceeds the maximum length limit of BERT, so that the document is divided into sentences or passages. Splitting the document and then aggregating the relevance scores increases the training and inference time.

### 9.2 Choice of embedding: how to compute query and document representations?

In addition to semantic and relevance matching signals, the context-sensitive embedding was shown to have better retrieval results than traditional pretrained embeddings like

Glove. A part of using context-sensitive embedding is to incorporate the query context into the ranking model to improve the precision of ad-hoc retrieval. Recent neural ranking models use deep contextualized pre-trained language models to compute a contextual representation for each token. There are mainly two advantages from using these models; first, they are bidirectional language representations, in contrast to only left-to-right or right-to-left language models so that every token can attend to previous and next tokens to incorporate the context from both directions. Second, they contain the attention mechanism which becomes an important component of sequence representations in multiple tasks, especially the Transformer's self-attention which captures long-range dependencies better than the recurrent architectures (Vaswani et al., 2017). So, this contextual representation covers both the context-aware representation and the attention-based representation. The expensive computation is still a limitation when incorporating the pre-trained language models. For example, the large BERT model has 340$M$ parameters consisting of 24 layers of Transformer blocks, 16 self-attention heads per layer and a hidden size of 1024. Zhan et al. (2020a) analyzed the performance of BERT in document retrieval using the model proposed by Nogueira and Cho (2019). The analysis showed that the [CLS], [SEP], and periods are distributed with a large proportion of attention because they appear in all inputs. BERT has multiple layers, and the authors showed that there is different behavior in different layers. The first layers are representation-focused, and extract representations for a query and a document. The intermediate layers learn contextual representations using the interaction signals between the query and document. Finally, for the last layers, the relevance score is predicted based on the interactions between the high-level representations of a given query-document pair.

External knowledge bases and graphs are incorporated into neural ranking models to provide additional embeddings for the query and document. Knowledge graphs contain human knowledge and can be used in neural ranking models to better understand queries and documents. In general, the entity-based representation, that is computed from knowledge bases, is combined with the word-based representation. In addition, the knowledge graph semantics, such as the description and type of entity, provide additional signals that can be incorporated into the neural ranking model to improve retrieval results and generalize to multiple scenarios.

The existing methods, that incorporate knowledge graphs into the ranking models, leverage mainly entities and knowledge graph semantics. However, knowledge graphs often provide rich axiomatic knowledge that can be explored in future research to improve retrieval results.

### 9.3 How to reduce the complexity of the neural ranking architectures that use deep contextualized language models?

After summarizing multiple existing neural ranking models, we can conclude that a potential future research direction is the design of more efficient neural ranking models that are able to incorporate semantic and relevance matching signals. Using BERT as a semantic matching component comes with the disadvantage of BERT length limit where the number of allowed tokens in BERT is significantly smaller than the typical length of a document. So, a possible future research direction is the study of selection techniques for sentences or passages with a trade-off between high retrieval results and low computation time. Ranking documents of length $m$ using Transformers, which are the main components of BERT, requires $O(m^2)$ memory and time complexity (Kitaev et al., 2020). In particular, for very long documents, applying self-attention of Transformers is not feasible. So, BERT-based ranking models have a large increase in computational cost and memory complexity over the existing traditional and neural ranking models. A current research direction is the design of efficient and effective deep language model-based ranking architectures. For example, Khattab and Zaharia (2020) presented a ranking model that is based on contextualized late interaction over BERT (ColBERT). The proposed model reduces computation time by extracting BERT-based document representations offline, and delays the interaction between query and document representations. RepBERT (Zhan et al., 2020b) and RocketQA (Qu et al., 2021) are other models proposed to solve the passage retrieval task following the same direction of designing a representation-based model with BERT being the main component to map the query and document. As in ColBERT, the objective is to overcome the computation time and memory limitations of the cross-encoding attentions related to the sentence pair setting of BERT. Despite the promising results of the BERT variants (ColBERT, RepBERT, RocketQA), reducing the memory complexity is still an open research question for multiple reasons. First, the document embeddings need to be loaded into the system or GPU memory (Johnson et al., 2021) with a limited size to compute the relevance scores of query-document pairs. Second, the dimension of the embedding is very large compared to the bag-of-word index (Zhan et al., 2021a; Xiong et al., 2021). Therefore, vector compression methods (Jégou et al., 2011; Ge et al., 2014) have been integrated into neural ranking models to compress the embedding index and save computational resources with compressed embeddings of documents. The compression methods include Product Quantization (PQ) (Jégou et al., 2011, 2014) and Locality Sensitive Hashing (LSH) (Indyk & Motwani, 1998). Improving the memory efficiency using index compression can lead to a drop in the performance of the neural ranking model. Zhan et al., (2021b) proposed a joint optimization of query encoding and PQ in order to maintain effectiveness of neural ranking models while compressing index sizes. The authors showed that an end-to-end training strategy of the encoding and compression steps overcomes the training based on the reconstruction error for many compression methods (Jégou et al., 2011; Ge et al., 2014; Guo et al., 2020).

In the same direction of reducing the ranking model complexity, Hofstätter et al. (2020) reduced the time and memory complexity of Transformers by considering the local self-attention where a given token can only attend to tokens in the same sliding window. In the particular case of non-overlapping sliding windows of size $w << m$, the time and memory complexity is reduced from $O(m^2)$ to $O(m \times w)$. Recently, Kitaev et al. (2020) improved the efficiency of Transformers and proposed the Reformer which is efficient in terms of memory and runs faster for long sequences by reducing the complexity from $O(m^2)$ to $O(m \times log(m))$. The Reformer presents new opportunities to achieve the trade-off between high retrieval results and low computation time. ElMo provides deep contextualized embeddings without the length limit constraint and can be used as a semantic matching component as it is pre-trained on the next word prediction. A possible multi-stage ranking architecture, that has a trade-off between retrieval results and computation time, can be composed of a first stage that re-ranks a set of candidate documents obtained from BM25 using an ElMo-based semantic and relevance model (example of relevance models: K-NRM, Conv-KNRM, DRMM, etc). Then, for the second stage, the top ranked documents from the first stage are re-ranked using a BERT-based semantic and relevance model. This multi-stage model has the potential to reduce the number of documents that should be ranked with BERT.

## 10 Summary

An end-to-end LTR architecture for ad-hoc document retrieval consists of extracting features from a query-document pair using a feature extractor, and mapping the extracted features to a real-valued relevance score using a ranking function. A core component of such a system is a neural ranking model, and in this survey, we have presented several such models. We began by introducing deep learning terminologies and techniques in Sect. 2 in order to cover the major components that are used in different neural ranking models. Then, We present the document retrieval field in Sect. 3. The objective of the document ranking and retrieval is to retrieve a set of documents where the top ranked documents are more relevant to the user's query.

After formulating the document ranking task using the LTR framework in Sect. 4, we reviewed the three learning to rank approaches used in ad-hoc retrieval in Sect. 5 namely, pointwise, pairwise, and listwise approaches. We summarized several techniques from each to address the advantages and drawbacks of each approach. In Sect. 6, we summarized the neural ranking models in ad-hoc retrieval, and we proposed placing the neural ranking models into five groups, where each group has unique neural components and architectures. In Sect. 7, we discussed the features of neural ranking models. In particular, we proposed nine features that are frequently presented in neural ranking models. These features are used for an in-depth overview of several neural ranking models that are proposed in the literature. This overview gives a global view for the current state of neural ranking models in addition to the work that has been done during the last several years. In Sect. 8, we showed that neural ranking models can generalize beyond the document retrieval. In particular, we described four related retrieval tasks: structured document retrieval, question-answering, image retrieval, and ad-hoc video search. Finally, in Sect. 9, we discussed the important signals in neural ranking models which are the semantic and relevance matching. In addition, we focused on a research direction that studies the trade-off between high quality retrieval results and fast computation mainly for ranking models with deep contextualized language models.

By grouping the neural ranking models, we analyzed the main features of each proposed model, and pointed to future research directions based on the important signals of the ad-hoc retrieval. In particular, we reviewed the design principles of neural ranking models that are used to extract the representation of a given query-document pair. The number of proposed neural ranking architectures is increasing, and this survey can help researchers by providing an overview about the current progress, and suggesting some potential directions to improve neural ranking models for document retrieval in both retrieval results and computation time.

**Author contributions** The first draft of the manuscript was written by all authors. All authors read and approved the final manuscript.

**Funding** This material is based upon work supported by the National Science Foundation under Grant No. IIS-1816325.

**Data availability and material** Not applicable.

**Code availability** Not applicable.

## Declarations

**Conflict of interest** Not applicable.